# Magnetic Proximity-Induced Superconducting Diode Effect and Infinite Magnetoresistance in van der Waals Heterostructure

Jonginn Yun[1†], Suhan Son[1,2†], Jeacheol Shin[1†], Giung Park, Kaixuan Zhang[1,2], Young Jae Shin[3], Je-Geun Park[1,2*], and Dohun Kim[1*]

[1] *Department of Physics and Astronomy, and Institute of Applied Physics, Seoul National University, Seoul 08826, Korea*

[2] *Center for Quantum Materials, Seoul National University, Seoul, 08826, Korea*

[3] *SC devices, PsiQuantum, Palo Alto, California, 94304, USA*

[†]*These authors contributed equally to this work.*

*Corresponding author: jgpark10@snu.ac.kr, dohunkim@snu.ac.kr*

**Abstract**

We report unidirectional charge transport in a $NbSe_2$ non-centrosymmetric superconductor, which is exchange-coupled with a $CrPS_4$ van der Waals layered antiferromagnetic insulator. The $NbSe_2/CrPS_4$ bilayer device exhibits bias-dependent superconducting critical-current variations of up to 16%, with the magnetochiral anisotropy reaching $\sim 10^5\ T^{-1}A^{-1}$. Furthermore, the $CrPS_4/NbSe_2/CrPS_4$ spin-valve structure exhibits the superconducting diode effect with critical-current variations of up to 40%. We also utilize the magnetic proximity effect to induce switching in the superconducting state of the spin-valve structure. Finally, it exhibits an infinite magnetoresistance ratio depending on the field sweep direction and magnetization configuration. Our result demonstrates a novel route for enhancing the nonreciprocal response in the weak external field regime (<50 mT) by exploiting the magnetic proximity effect.

## I. INTRODUCTION

Nonreciprocal charge transport, where the electrical resistivity varies depending on the current and magnetic-field direction, has attracted considerable attention over the past decade. Extensive experimental and theoretical research [1–12] has been conducted on systems with broken inversion or time-reversal symmetries. With notable recent progress, including the nonlinear Hall effect in Weyl semimetals [13,14] and chiral optical and electrical responses in semiconductors [2,15–17], new efforts have been directed toward exploiting nonreciprocal transport in search of novel physical properties and functionalities.

In a system with both broken inversion and time-reversal symmetries, the nonlinear resistance $R$, in general, quantify the nonreciprocal transport, which is phenomenologically expressed as [1–3]

$$R = R_0(1+\gamma IB), \quad (1),$$

where $B$ represents the external magnetic field, and $I$ denotes the electric current. The second term represents nonreciprocal transport, which depends on the directions of both $I$ and $B$. The coupling coefficient $\gamma$, which is known as magnetochiral anisotropy (MCA), defines the strength of the aforementioned dependence. Typical metals exhibit small MCA values ranging from $10^{-3}$ to $10^0$ $T^{-1}A^{-1}$ because the spin-orbit interaction and magnetic energies are nominally far lower than the kinetic energy of the electrons, typically within a few electronvolts of the Fermi energy $E_F$ [1,2,7,10]. However, measuring the MCA in non-centrosymmetric superconducting materials such as transition-metal dichalcogenides (TMD) is a promising route for enhancing the MCA by orders of magnitude ($\sim 10^4$ $T^{-1}A^{-1}$ [4–6]) compared with the normal state, by replacing the role of $E_F$ by superconducting gap [4,7]. Therefore, nonreciprocal charge transport is potentially suitable for developing phase-coherent and direction-selective electronic devices, and a proof-of-principle demonstration has been performed for a broadband antenna [5].

The nonreciprocity quantified by the MCA generally arises from finite inequivalent resistances under different transport directions. In contrast, exotic nonreciprocity can be achieved if the critical current becomes nonreciprocal. In such a case, the resistance is quenched for one direction while remaining finite for the other. If successful, this exotic nonreciprocity can be used to develop dissipation-less electric circuits [18]. The recent discovery of such a superconducting diode effect (SDE) in a Nb/V/Ta epitaxial film has prompted research on the nonreciprocal transport in superconductors [8]. In subsequent experiments, the SDE was observed in non-centrosymmetric superconductors [19], artificially nanofabricated systems such as a Josephson junction [20], and a non-centrosymmetric superconducting film with conformal-mapped nanoholes [21].

In general, nonreciprocal transport is observed using an external magnetic field for breaking time-reversal symmetry. Several studies have been performed on enhancing the MCA near the superconducting-normal state transition [22]. However, the commonly adopted approach typically requires a large magnetic field, on the order of a few teslas [4–6], which poses a challenge for practical application. This technical bottleneck provides a solid impetus to developing an alternative method for facilitating time-reversal symmetry breaking.

The proximity effect in superconductor and ferromagnet (S/F) heterostructures offers a possible solution to a large magnetic field requirement. In the case of a superconductor in close contact with a ferromagnet, the magnetic exchange field of the ferromagnet is expected to penetrate into the superconducting layer [23,24]. For example, the exchange field in the S-ferromagnet insulator (FI) heterostructure has been adequately understood based on this inverse proximity effect [25,26]. Exchange fields of a few teslas have often been observed in the spin splitting of the quasiparticle density of states in EuS/Al bilayers [27–29] and EuO/Al bilayers [30]. This large exchange field paves the way for developing novel superconducting spintronic devices. For instance, superconducting spin-valve structures exhibiting infinite magnetoresistance (MR) have been recently fabricated [31,32].

Here, we show that when a magnetic proximity-coupled van der Waals (vdW) heterostructure is used, the exchange field effectively reduces the otherwise large external magnetic field requirement while preserving or even enhancing the MCA in a non-centrosymmetric superconductor. Using a recently developed polymer-based strong adhesive transfer technique [33], we manufactured a NbSe$_2$/CrPS$_4$ bilayer heterostructure and a CrPS$_4$/NbSe$_2$/CrPS$_4$ trilayer spin-valve structure to leverage the magnetic proximity effect of the adjacent CrPS$_4$ layer—an A-type layered antiferromagnetic insulator (AFI) below $B \approx 0.7$ T [34,35]. The key observation was that a nonreciprocal SDE with a critical-current variation of up to 16% under a small external magnetic field of <50 mT in the bilayer device and up to 40% in the trilayer device. Moreover, an infinite MR was observed in the trilayer device under specific probe currents, which had never been demonstrated in a vdW heterostructure or S/AFI heterostructure.

## II. METHODS

### A. Device fabrication

The NbSe$_2$/CrPS$_4$ bilayer and CrPS$_4$/NbSe$_2$/CrPS$_4$ trilayer devices are fabricated using the polycaprolactone (PCL) dry-transfer method [33]. Specifically, the exfoliated layers of NbSe$_2$ were picked up by the PCL stamp and dropped onto the layers of CrPS$_4$ to minimize the polymer residue at the interface. Then, Ti (5 nm)/Au (70 nm) electrodes were evaporated after the conventional e-beam lithography.

### B. Transport measurements

The transport measurement was performed using a commercial variable temperature cryostat (Teslatron PT, Oxford Instruments) with a base temperature of $T = 1.5$ K. We performed two different types of four-probe transport measurements in this study. The first was total resistance ($R$ in Eqn. (1)) measurement using the direct-current voltage $V$ with a Keithley 2182A Nanovoltmeter. The second measurement method involved using the first and second harmonic alternating-current resistances ($R_0$ in Eq. (1) and nonlinear correction to $R_0$ due to nonreciprocity, respectively) with a standard lock-in amplification technique (Signal Recovery, 7265). Finally, the device was field-cooled at $B = 8$ T to increase the domain size of the CrPS$_4$ layers [27,29,36–38].

### III. RESULTS

#### A. Nonreciprocal response in the bilayer device

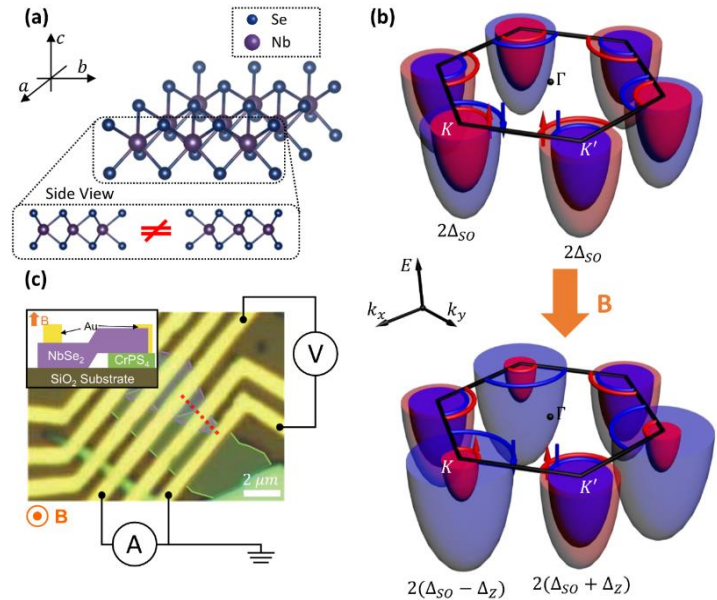

**Figure 1.** (a) Structure of non-centrosymmetric NbSe$_2$. Inset: Side view of the structure of NbSe$_2$, where the chirality is illustrated. (b) Schematic of the valley-dependent splitting (top) and the spin-valley locking of NbSe$_2$ (bottom). The broken inversion symmetry of NbSe$_2$

introduces spin-orbit coupling, which causes valley-dependent spin-orbit splitting ($\Delta_{SO}$). The Zeeman splitting ($\Delta_Z$) due to the downward magnetic field causes spin-valley locking. The Fermi surfaces of the spin branches are plotted with different colors. **(c)** Optical microscope image of the device. The boundaries of NbSe$_2$ and CrPS$_4$ flakes are marked with violet and green lines, respectively. Beneath the NbSe$_2$ were layers of antiferromagnetic CrPS$_4$, which were placed on top of the 285-nm-thick SiO$_2$ substrate. The Au electrodes were placed on top of the NbSe$_2$, and a four-probe measurement was conducted. The applied magnetic field was perpendicular to the substrate. The inset shows a schematic of the cross-section marked with the red dotted line.

As shown in Fig. 1a, the lattice structure of monolayer 2*H*-NbSe$_2$ possesses intrinsically broken in-plane inversion symmetry originating from inequivalent Nb and Se sites. Consequently, the itinerant electron of NbSe$_2$ is subjected to an effective out-of-plane magnetic field through Ising-type spin-orbit coupling. This effective magnetic field causes the spin-dependent valley splitting of NbSe$_2$ in Fig. 1b, as theoretically predicted [39] and experimentally confirmed using photoemission spectroscopy [40,41]. With locally broken inversion symmetry and spin-valley locking, the nonreciprocal behavior is expected to occur when the time-reversal symmetry is broken by a magnetic field perpendicular to the plane.

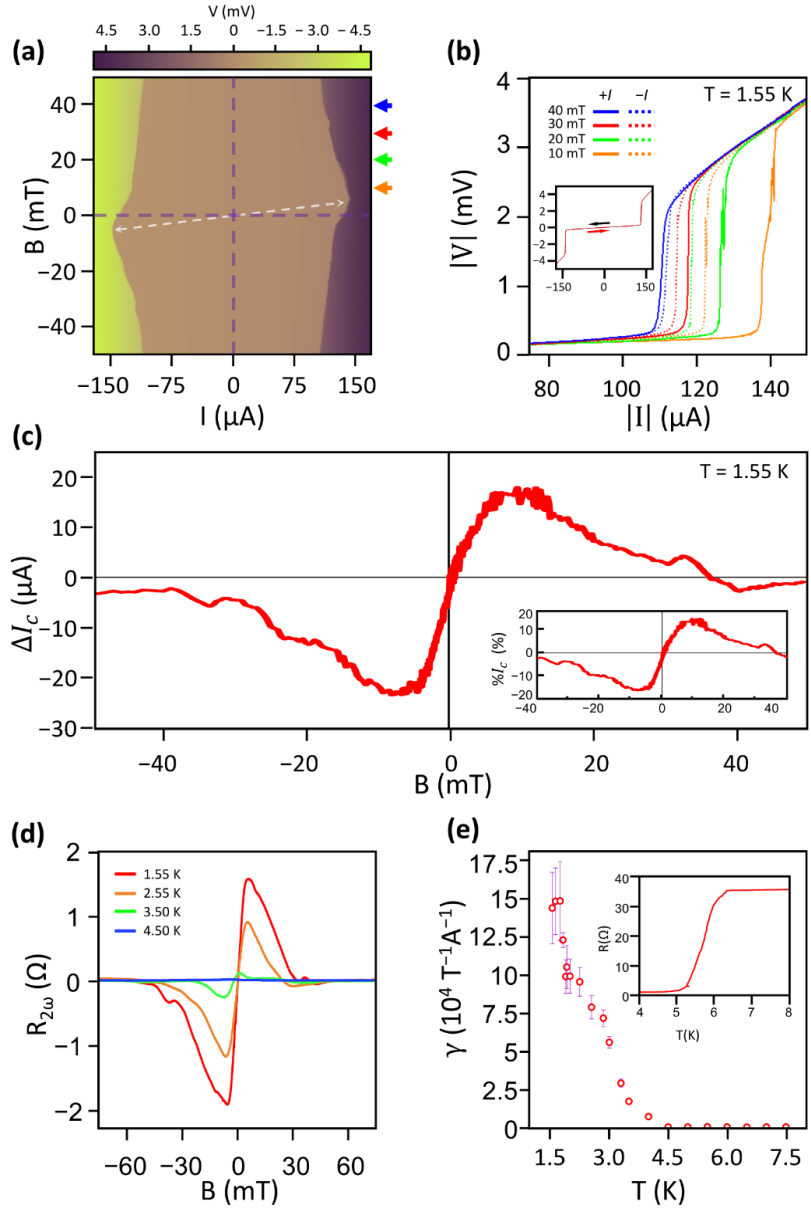

**Figure 2. (a)** Voltage drop across the NbSe$_2$ layer with respect to the current and magnetic field at $T = 1.55\,\mathrm{K}$. The nonreciprocity and magnetochirality appear as point symmetry, as indicated by the purple dotted line. **(b)** *I-V* line cuts of Fig. 3a at 10, 20, 30, and 40 mT. The solid and dotted lines indicate the positive and negative transports direction, respectively. The line cuts are plotted near $I_c$ to highlight the nonreciprocal $I_c$. The location of each curve in the colormap of Fig. 3a is marked with arrows of the same color. The inset shows the current-voltage characteristic with a different current sweep direction at $B = 0\,\mathrm{T}$, where the arrows

with the corresponding color indicate the current sweep direction. There is no noticeable change in $I_c$, which suggests that the observed SDE is not by a retrapping current. **(c)** Dependence of $\Delta I_c$ on the applied magnetic field. The measurement was performed by sweeping the magnetic field from positive to negative. The inset shows the dependence of $\%I_c$ on the applied magnetic field. A maximum value of 16% was observed. **(d)** Magnetic-field dependence of the second-harmonic resistance. The magnetic field was swept from positive to negative. The temperature of each trace is indicated in the upper left legend. Magnetochirality was observed, and the peak location shifted to lower *B* values as the temperature increased. **(e)** MCA $\gamma$ was calculated from Fig. 2d using Eq. (2) with the corresponding error bar. The inset shows the temperature dependence of the resistance of the device.

We first investigate the nonreciprocal response in the bilayer device shown in Fig. 1c, which is fabricated using the PCL stamp technique (see Methods). Fig. 2a shows the *I-V* characteristics of NbSe$_2$ as a function of *B*. The distribution of the superconducting critical current $I_c$ across the four quadrants of the colormap reveals broken symmetry in both the *I* and *B* directions. Fig. 2b shows the line cuts of the magnitudes of the *I-V* curve in the vicinity of $I_c$ under different *B* values, introducing a clear SDE observed in the bilayer structure. Notably, the $I_c$ does not exhibit a significant change even when the current sweep direction is varied (inset of Fig. 2b), suggesting the effect of the retrapping current that an accidental Josephson junction could have caused can be disregarded. The complete field dependence of the marked critical-current difference $\Delta I_c$ is plotted in Fig. 2c, which is defined as the difference in $I_c$ between the positive ($I_{c,+}$) and negative ($I_{c,-}$) current-bias directions. The point symmetry of nonzero $\Delta I_c$ exhibits clear magnetochirality. Furthermore, $\%I_c$, which is defined as $\%I_c = \frac{\Delta I_c}{\min(I_{c,+}, I_{c,-})} \times 100$, reaches a value of up to 16% (inset of Fig. 2c).

The second-harmonic resistance $R_{2\omega}$ is shown as a function of $B$ at different $T$ values in Fig. 2d. The $B$ field where the largest $R_{2\omega}$ occurs generally coincides with that of $\Delta I_c$. The MCA can then be quantified using $R_{2\omega}$, as follows [4]:

$$\gamma = \frac{2R_{2\omega}}{R_0 BI}, \quad (2),$$

where the external field $B$ is used to estimate $\gamma$ solely from experimentally given conditions. Interestingly, Fig. 2e shows a $\gamma$ value (MCA) of $1.5 \times 10^5$ $T^{-1}A^{-1}$ at $T = 1.5$ K, mainly because of the reduced $B$ field compared with that observed in previous studies [4–6]. A monotonic decrease in $\gamma$ as a function of $T$ is observed before $\gamma$ becomes negligible at approximately 5 K, below the superconducting critical temperature $T_c = 5.51$ K (inset of Fig. 2e) [42,43].

The observed $R_{2\omega}$ is different from that observed in the few-layer NbSe$_2$ of the previous study [5]. Previously, nonzero $R_{2\omega}$ in NbSe$_2$, which was attributed to the resistive vortex flow regime, started to appear at $B > 1$ T for $T < 4$ K [5]. However, in our case, nonzero $R_{2\omega}$ appears below $B < 0.06$ T for $T < 4$ K. We ascribe the unusual behavior of $R_{2\omega}$ to the magnetic ordering of CrPS$_4$. As mentioned, A-type AFI CrPS$_4$ layers consist of antiferromagnetically coupled ferromagnetic monolayers for $B < 0.7$ T at $T = 1.5$ K [34,35]. Furthermore, because the proximity effect in the S/FI insulator is confined to the interface by the exponential decay of the electronic wave function in the insulator, the exchange coupling between the ferromagnetic CrPS$_4$ monolayer and NbSe$_2$ is expected to decay rapidly within the atomic distance [23,25,31]. Therefore, the antiferromagnetically coupled upper layers do not effectively offset the exchange field provided by the nearest ferromagnetic CrPS$_4$ monolayer, resulting in a significant exchange field dominated by the magnetization of the nearest monolayer. This exchange field alone does not affect the orbital dynamics because the exchange field couples only with the quasiparticle spin and does not induce a Lorentz force.

Additionally, the antiferromagnetic nature of $CrPS_4$ results in a negligible stray field [44]. Thus, the stray field of $CrPS_4$ and the applied magnetic field do not provide the net magnetic field required to achieve the vortex flow regime, which is $B > 2$ T for a few-layer $NbSe_2$ at a temperature below 2 K [5]. From this perspective, our observation of $R_{2\omega}$ differs significantly from previously reported vortex dynamics-based $R_{2\omega}$ on $NbSe_2$.

## B. Nonreciprocal response in the trilayer device

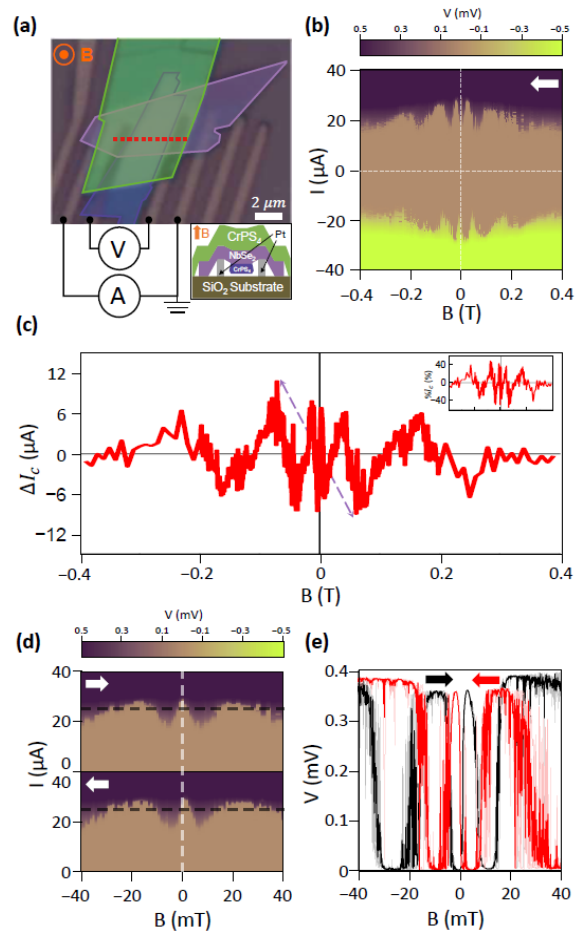

**Figure 3. (a)** Optical microscope image of the trilayer spin-valve device. The boundaries of NbSe$_2$ and the top (bottom) CrPS$_4$ flakes are marked with violet and green (purple) lines, respectively. Two antiferromagnetic CrPS$_4$ layers sandwiched the NbSe$_2$ layer, and the entire devices were placed on top of the 285-nm-thick SiO$_2$ substrate. The Au electrodes were at the bottom of the device, and a four-probe measurement was conducted. The applied magnetic field was perpendicular to the substrate. The inset shows a cross-sectional schematic illustrating the stacking sequence. **(b)** Colormap of the four-probe voltage with respect to the magnetic field and the current. The magnetic field was swept from positive to negative. A drastic change in $I_c$ near $B = 0$ T was observed. **(c)** Dependence of $\Delta I_c$ on the applied magnetic field. The inset shows the dependence of $\% I_c$ on the applied magnetic field. A maximum value of ~40%

was observed. **(d)** Magnified colormap of the four-probe voltage near $B=0\,\text{T}$, where the magnetic field was swept from negative to positive (top) and from positive to negative (bottom). A significant change in $I_c$ is clearly observed. This change produced a large MR under a well-chosen probe current, such as $I=25\,\mu\text{A}$ (black dotted line). Hysteresis was also observed, which reflects the effect of CrPS$_4$. **(e)** Four-point voltage as a function of the magnetic field measured with a current magnitude of $I=25\,\mu\text{A}$, where the magnetic field was swept from positive to negative (red) and from negative to positive (black). The averaged signals of five traces are plotted, with each trace collectively plotted in the background with lighter colors. The infinite MR and the hysteresis are clearly observed in the averaged curve. Stochastic switching also appears in each trace plotted in the background.

We next explore the response of the similar proximity-coupled structure with the increased exchange field by CrPS$_4$/NbSe$_2$/CrPS$_4$ trilayer spin valve structure shown in Fig. 3a. The device is fabricated using the previously mentioned PCL stamp technique, where the picked-up material stack was dropped directly onto pre-patterned Pt(18 nm)/Ti(2 nm) electrodes to avoid contamination.

Fig. 3b shows the current-voltage (*I-V*) characteristics of the device as a function of $B$, whose sweep direction is indicated in the upper-right corner. The phase boundary in the *I-B* plane manifests both magnetochirality and nonreciprocity, as highlighted by the white dotted lines. We quantify the observed SDE in the trilayer structure by plotting $\Delta I_c$ and $\%I_c$ in Fig. 3c. Notably, $\%I_c$ again shows a substantial value with the increased value reaching 40% (inset of Fig. 3c).

Furthermore, a significant magnetic dependence of $I_c$ near $B = 0$ T was observed, as shown in the enlarged plot of Fig. 3d. This dependence is also revealed in the averaged four-probe voltage measured by sweeping the magnetic field at $I = 25$ μA in Fig. 3e, with the noisy stochastic fluctuating behavior depicted in the background of the averaged signal. This stochastic fluctuation may be related to the phase slip line (PSL) and kinematic vortices [45–47], the existence of the metastable state near the superconductor-normal state transition [48], or the domain structure of the proximity-coupled $CrPS_4$ and their dynamics to be discussed below. Nonetheless, the overall signal consistently showed the field-dependent switching behavior, as shown in the averaged profile, demonstrating an infinite MR.

We note that the observed $I_c$ change near $B = 0$ T is related to the relative alignment of the adjacent $CrPS_4$ layer. As mentioned previously, $CrPS_4$ provides an effective exchange field with the dominant effect of the nearest ferromagnetic layer. Thus, the system can be effectively identified with a FI/S/FI trilayer structure, and the average exchange field on the quasiparticles is defined as follows:

$$\bar{h} = 2|\Gamma|S(a/d_s)\cos(\theta/2), \quad (3),$$

where $|\Gamma|S$ represents the ferromagnetic spin multiplied by the coupling constant, $\theta$ represents the angle between the magnetizations of the two ferromagnetic layers, and $a$ and $d_s$ represent the lattice constant and the thickness of the superconductor, respectively [25]. In our device, the coercive field of the nearest $CrPS_4$ layers is tuned according to their thicknesses. This condition allows the switching of the device between parallel (P) and antiparallel (AP) configurations through a gradual reduction in the magnetic field. Thus, in the P configuration ($\theta = 0°$), the net exchange field between two $CrPS_4$ layers results in a significant exchange field on $NbSe_2$.

In contrast, in the AP configuration ($\theta = 180°$), the net exchange field on NbSe$_2$ is canceled out [31,32]. This magnetization reversal of the nearest CrPS$_4$ layers considerably changes the exchange field on NbSe$_2$ near $B = 0\,\text{T}$. Considering that the exchange field tends to reduce $I_c$ because it disrupts the Cooper pairs, the analogous switching behavior of $I_c$ near $B = 0\,\text{T}$ reveals the exchange coupling between the NbSe$_2$ and CrPS$_4$ layers, along with the hysteresis appearing in the colormap of Fig. 3d. This hysteresis becomes more apparent when the field is swept with the fixed current as in Fig. 3e, which could be distinguished from the stochastic fluctuation.

## IV. DISCUSSION

The bilayer device shows no significant difference from the pure NbSe$_2$ device except for the five-fold decrease of the magnetic field for the maximal %$I_c$ and the unusual behavior of $R_{2\omega}$ [19]. However, the trilayer device shows several exotic features, including infinite MR near $B = 0\,\text{T}$ that we regard as a fingerprint of the magnetic proximity effect. Another novel feature of the trilayer device is the fluctuation of $I_c$ in the colormap of Fig. 3b. If we interpret this fluctuation as a Fraunhofer-like aperiodic pattern of a Josephson junction, the weak link must have a width of ~10 nm for the ~4-μm-long device. Since there are no reasonable candidates for such a weak link, the fluctuation of $I_c$ most plausibly originates from the fluctuation of the magnetization of CrPS$_4$. In addition, we observe that the $I_c$ fluctuation vanishes $B \approx 0.7\,\text{T}$, which is close to the phase transition of CrPS$_4$ between the A-type antiferromagnetic phase and the canted antiferromagnetic phase [35,49]. This vanish of the fluctuation also reveals the connection between the $I_c$ and the magnetization of CrPS$_4$.

However, the exact reason for the magnetization fluctuation remains unclear. We suspect the fluctuation is possibly related to the layer-by-layer switching behavior, as observed in certain layered antiferromagnets [36,37] and ferromagnets [50]. Further research on the magnetic structure of CrPS$_4$ is required to elucidate the underlying principle for the fluctuation of $I_c$, and we leave it for future work.

We next discuss the SDE observed in both devices. As previously noted, the effect of CrPS$_4$ on the $I_c$ of the trilayer structure is revealed by the $I_c$ fluctuation. Furthermore, we also observed that the SDE in NbSe$_2$ disappears when the nearest CrPS$_4$ layers are not present in another bilayer structure [49]. These observations suggest that the observed SDE in the NbSe$_2$/CrPS$_4$ heterostructure is most likely induced by the magnetic proximity effect of the adjacent CrPS$_4$ layer.

Although theoretical analyses on the SDE observed in Junction-free superconductors are still in their infancy, recent theories propose finite-momentum Cooper pairing (FMCP) as the microscopic origin of intrinsic SDE. Therefore, it is worth noting that the %$I_c$ of the trilayer device exhibits an abrupt increase near $I_c$ and saturates far below $T_c$, in line with the theoretically expected formula $\sqrt{1 - T/T_c}$ based on the FMCP scenario [49,51,52]. Additionally, the theory predicts the %$I_c$ of a magnetic proximity-coupled Ising superconductor to be proportional to the magnitude of the exchange field [49]. Considering that the effective exchange field of the trilayer structure in the P configuration is expected to be twice that of the bilayer structure, the observed increase of %$I_c$ in the trilayer device in this work compared to the bilayer device is seemingly consistent with the FMCP-based theory. However, it should be noted that the %$I_c$ observed in another device falls short of half of that of the trilayer structure, and that the parameters, including layer numbers and transport direction, are not controlled in the experiments. Moreover, a recent study that observed SDE in pure NbSe$_2$ reports that %$I_c$ shows uncorrelated random behavior regardless of the layer number and supercurrent-to-lattice orientation [19]. Thus, further experimental and theoretical investigations are required to fully comprehend the underlying microscopic origin behind the observed behavior.

In conclusion, we demonstrated that the magnetic proximity effect in the NbSe$_2$/CrPS$_4$ heterostructure provides a large exchange field to induce clear nonreciprocal transport. The NbSe$_2$/CrPS$_4$ bilayer device exhibited the SDE, where the magnetochiral $\Delta I_c$ and $R_{2\omega}$ were observed for $B < 50$ mT, and $\gamma$ reached $\sim 10^5$ T$^{-1}$A$^{-1}$. We also observed an infinite MR and SDE in the CrPS$_4$/NbSe$_2$/CrPS$_4$ spin-valve structure, with %$I_c$ reaching ~40%. Finally, we expect the time-reversal symmetry breaking caused by the magnetic proximity effect to provide a novel route for investigating the potential application of nonreciprocal transport under small (or even zero) external magnetic fields.

**Acknowledgments**

We thank Philip Kim and Gilho Lee for their valuable discussions. The National Research Foundation of Korea supported this work (NRF) grants funded by the Korean Government (MSIT) (Nos. 2018R1A2A3075438, 2019M3E4A1080144, 2019M3E4A1080145, and 2019R1A5A1027055), Korea Basic Science Institute (National Research Facilities and Equipment Center) grant funded by the Ministry of Education (No.2021R1A6C101B418), the Creative-Pioneering Researchers Program through Seoul National University (SNU), and the Leading Researchers Program of the National Research Foundation of Korea (No. 2020R1A3B2079375).

# Magnetic Proximity-Induced Superconducting Diode Effect in van der Waals Heterostructure

Jonginn Yun[1†], Suhan Son[1,2†], Jeacheol Shin[1†], Giung Park, Kaixuan Zhang[1,2], Young Jae Shin[3], Je-Geun Park[1,2*], and Dohun Kim[1*]

[1] Department of Physics and Astronomy, and Institute of Applied Physics, Seoul National University, Seoul 08826, Korea

[2] Center for Quantum Materials, Seoul National University, Seoul, 08826, Korea

[3] SC devices, PsiQuantum, Palo Alto, California, 94304, USA

[†]These authors contributed equally to this work.

*Corresponding author: jgpark10@snu.ac.kr, dohunkim@snu.ac.kr

**Supplemental Figures**

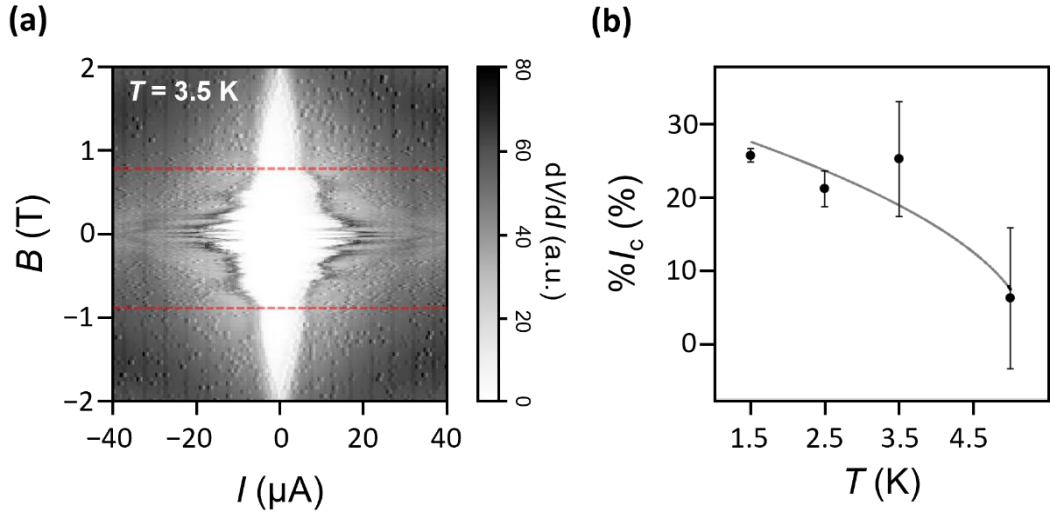

**Figure S1. (a)** A colormap of the d$V$/d$I$ of the trilayer device as a function of the current and magnetic field at $T = 3.5$ K. Note that the $I_c$ fluctuation disappears near $B \approx \pm 0.7$ T, which are marked by the red dotted lines. **(b)** Critical current efficiency %$I_c$ at $T$ = 1.5, 2.5, 3.5, and 5 K under the fixed magnetic field $B$ = 2 mT. The solid curve indicates a theoretically expected formula $A\sqrt{1 - T/T_c}$ from Ref. [1,2] that is fit to the observed data. The error bars are determined from the uncertainty caused by the broadness of the d$V$/d$I$ signal near the superconductor-normal metal transition.

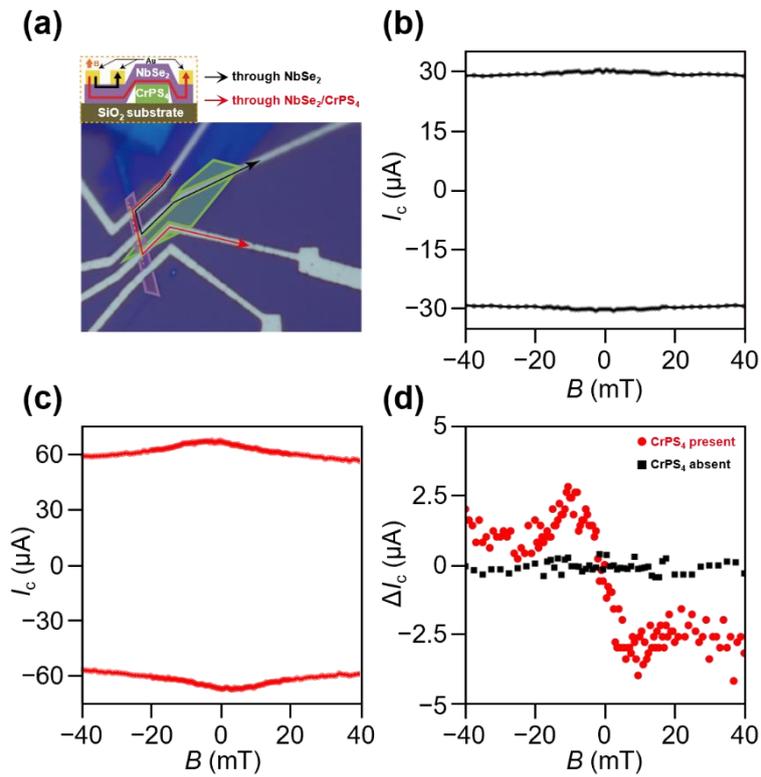

**Figure S2. (a)** An optical microscope image of the device, where the NbSe$_2$ and CrPS$_4$ layers are demarcated by purple and green lines, respectively. The red and black arrows indicate the current paths with and without the adjacent CrPS$_4$. **(b)** The critical current $I_c$ of the device when the current flows through NbSe$_2$ layers without CrPS$_4$. The current direction is indicated by the black arrow in Fig. S2(a). No superconducting diode effect is observed. **(c)** The $I_c$ of the device when the current flows through NbSe$_2$ in close contact with CrPS$_4$ layers. The current direction is indicated by the red arrow in Fig. S2(a). A superconducting diode effect is observed. **(d)** The critical current difference $\Delta I_c$ of the device when CrPS$_4$ is present (red circle) and absent (black square). The correlation between the CrPS$_4$ layer and the superconducting diode effect of NbSe$_2$ can be clearly identified.

**Supplemental Discussion**

Fig. S2(a) shows the optical microscope image of the device. The device has a region where the NbSe$_2$ layers lack the adjacent CrPS$_4$ layers, which enables the investigation of the superconducting diode effect (SDE) by controlling the existence of CrPS$_4$ layers. The $I_c$ of NbSe$_2$ with and without the adjacent CrPS$_4$ layers was measured by selecting the black and red arrow current paths in Fig. S2(a).

Fig. S2(b) and (c) present the resultant $I_c$ of the NbSe$_2$ layers with and without CrPS$_4$ layers, respectively. Notably, the $I_c$ does not show nonreciprocal behavior when the adjacent CrPS$_4$ is absent, whereas a clear SDE emerges when the adjacent CrPS$_4$ is present. The difference in the behavior of the $I_c$ of NbSe$_2$ depending on the presence of the CrPS$_4$ layer is more clearly illustrated in Fig. S2(d), which plots the critical current difference $\Delta I_c \equiv I_{c,+} - I_{c,-}$ of each current path. The device exhibits significant $\Delta I_c$ when the adjacent CrPS$_4$ is present, whereas no clear $\Delta I_c$ is observed when the CrPS$_4$ is absent. Overall, the results demonstrate that the CrPS$_4$ layers have a considerable impact on the nonreciprocity of the NbSe$_2$ device.